\documentclass[aps,twocolumn,superscriptaddress,prl,fleqn,showpacs,nofootinbib,preprintnumbers]{revtex4}

\usepackage{amssymb,amsmath,epsfig}
\usepackage{array}
\usepackage{color}

\DeclareMathOperator{\tr}{Tr}

\newcommand{\be}{\begin{equation}}
\newcommand{\ee}{\end{equation}}
\newcommand{\bea}{\begin{eqnarray}}
\newcommand{\eea}{\end{eqnarray}}

\newcommand{\st}{{\scriptscriptstyle T}}

\makeatletter
\newcommand{\thickline}{
	\noalign {\ifnum 0=`}\fi \hrule height 1pt
	\futurelet \reserved@a \@xhline
}
\newcolumntype{"}{@{\hskip\tabcolsep\vrule width 1pt\hskip\tabcolsep}}
\makeatother

\begin{document}

\title{Color entanglement for azimuthal asymmetries in the Drell-Yan process}

\author{M.G.A. Buffing}
\email{m.g.a.buffing@vu.nl}
\affiliation{Nikhef and Department of Physics and Astronomy, VU University Amsterdam,\\
De Boelelaan 1081, NL-1081 HV Amsterdam, the Netherlands}

\author{P.J. Mulders}
\email{mulders@few.vu.nl}
\affiliation{Nikhef and Department of Physics and Astronomy, VU University Amsterdam,\\
De Boelelaan 1081, NL-1081 HV Amsterdam, the Netherlands}

\begin{abstract}
In the resummation of collinear gluons emitted together with active partons from the hadrons in the Drell-Yan process (DY) effects of color entanglement become important when the transverse directions are taken into account. It is then no longer possible to write the cross section as the convolution of two soft correlators and a hard part. We show that the color entanglement introduces additional color factors that must be taken into account in the extraction of transverse momentum dependent parton distribution functions (TMD PDFs) from azimuthal asymmetries. Examples where such effects matter are the extraction of the double Sivers and double Boer-Mulders asymmetries. Furthermore, we will argue why this color entanglement is a basic ingredient already in the tree-level description of azimuthal asymmetries.
\end{abstract}

\date{\today}

\preprint{NIKHEF 2013-028}

\maketitle

\section{Introduction}
In Ref.~\cite{Rogers:2010dm} it was shown that the inclusion of contributions of collinear gluons in high-energy hadroproduction processes leads to the entanglement of color, complicating factorization of the cross sections into a hard part and soft correlators. Collinear gluons refer to gluons emitted from each of the target hadrons, with polarization along the hadron momentum. In Ref.~\cite{Buffing:2011mj} it was argued that this complication of factorization is even important at tree-level, where gauge links lead to color entanglement in the process, making it impossible to write a process with two initial state hadrons as the product of two correlators. These complications do not imply that observables can no longer be calculated, merely that results are different from the naive picture and have a richer phenomenology. In this paper, we will focus on the Drell-Yan process only~\cite{Drell:1970wh} and show in more detail what is different and how this affects measurements of asymmetries. We will use the results of Ref.~\cite{Buffing:2012sz} to discuss in general all asymmetries accessible through Drell-Yan involving unpolarized or polarized TMD PDFs at leading order in an expansion in $1/Q$, often sloppily referred to as `at leading twist'. We will also show why this effect of color entanglement is an essential ingredient, already at tree-level.

\section{Wilson lines at tree-level}
\begin{figure}[!b]
\centering
\epsfig{file=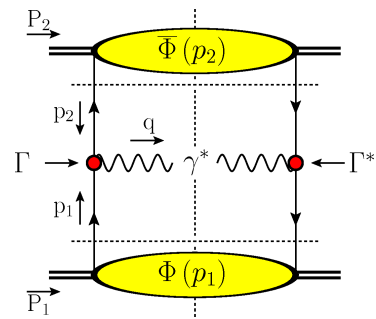,width=0.2663\textwidth}
\parbox{0.45\textwidth}{\caption{\label{f:DY}\textit{The DY process in the diagrammatic representation, where the yellow blobs are described by the TMD PDFs. The $\Gamma$ and $\Gamma^*$ represent the hard scattering, producing a virtual photon.}}}
\end{figure}

The leading order tree-level Drell-Yan cross section before taking into account gauge links, which are also leading order contributions, is illustrated in Fig.~\ref{f:DY} and given by
\bea
d\sigma_{\text{DY}}&\sim&
\tr_c\Big[\Phi(x_1,p_{1\st})\Gamma^{*}\overline\Phi(x_2,p_{2\st})\Gamma\Big]
\nonumber \\ &=&\frac{1}{N_c}\Phi(x_1,p_{1\st})\Gamma^{*}\overline\Phi(x_2,p_{2\st})\Gamma.
\label{e:basic-0}
\eea
Here $\Phi$ and $\overline\Phi$ are the TMD quark and antiquark correlators respectively, Fourier transforms of forward matrix elements of quark fields, and $\Gamma$ and $\Gamma^*$ represent the hard scattering interaction in which a virtual photon or weak vector boson with momentum $q$ is produced. The standard color factor emerges because the color trace is usually included in the definition of the correlator $\Phi$, i.e.\ $\tr_c[1]/(\tr_c[1]\,\tr_c[1]) = 1/N_c$. This is the basic expression of the TMD factorized parton model description after expanding the correlator into TMD PDFs. The result involves soft parts integrated over parton virtualities and is actually a convolution over the parton momenta $p_i = x_i\,P + p_{i\st}$. High-energy kinematics links the momentum fractions (or $p^+$ components) to scaling variables $x_1 = P_2{\cdot}q/P_1{\cdot}P_2$ and $x_2 = P_1{\cdot}q/P_1{\cdot}P_2$ and the sum of transverse momenta to the observable transverse momentum $p_{1\st}+p_{2\st} = q_\st \equiv q-x_1\,P_1-x_2\,P_2$, which is the transverse momentum of the virtual photon or the lepton pair with respect to the momenta $P_1$ and $P_2$, see Ref.~\cite{Ralston:1979ys}.

\begin{figure}[!bt]
\centering
\epsfig{file=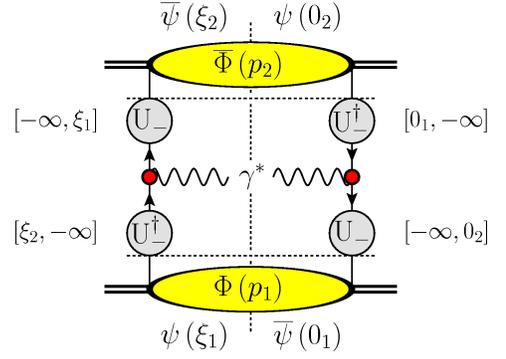,width=0.35\textwidth}
\parbox{0.45\textwidth}{\caption{\label{f:DYGL}\textit{The gauge connections contributing for Drell-Yan, indicated by gray blobs at the location in the diagram where they appear after resummation, the coordinates in brackets labelling the endpoints of the Wilson lines in coordinate space. The separations $\xi_i$ are conjugate to parton momenta $p_i$ involving light-cone $\xi^-$ and $\xi_\st$ directions. The $U_-$ gauge connections run to light-cone $\xi^- = -\infty$.}}}
\end{figure}

In the $q_\st$-integrated situation, the absorption of collinear gluons in the correlators $\Phi$ as color gauge links is simple. The correlators only depend on momentum fractions that are conjugate to light-like nonlocalities in the expressions in terms of partonic fields. Gauge links are just simple straight Wilson lines. At measured $q_\st$, determining the cross section for Drell-Yan includes gauge links with transverse separations involving collinear and transverse gluons. In the process the color remains entangled as illustrated in Fig.~\ref{f:DYGL}. Bypassing the details of getting gauge links in the first place, we note that at measured $q_\st$ the ingredients that contribute to the gauge links appear in different parts of the diagram and cannot be trivially absorbed in the definition of the TMD correlators, nor can they be incorporated by a simple redefinition of the correlator. Therefore, the name gauge connection rather than gauge link is used at this point. The result is
\bea
d\sigma_{\text{DY}}&=& \tr_c\Big[U_{-}^{\dagger}[p_2]\Phi(x_1,p_{1\st})U_{-}[p_2]\Gamma^{*} \nonumber \\
&&\hspace{5mm}\times U_{-}^{\dagger}[p_1]\overline\Phi(x_2,p_{2\st})U_{-}[p_1]\Gamma\Big]
\label{e:basic-2} \\
&\neq &\frac{1}{N_c}\,\Phi^{[-]}(x_1,p_{1\st})\Gamma^{*}\overline\Phi^{[-^\dagger]}(x_2,p_{2\st})\Gamma ,
\nonumber
\eea
suppressing all parts of the (partial) cross section that are not of direct importance for our purpose, e.g.\ the phase space factors. As arguments of the Wilson lines we have used a notation with the momenta $p_1$ and $p_2$ in square brackets, merely to indicate from which correlator the gauge connections receive contributions in the form of gluon emissions. The second expression in Eq.~\ref{e:basic-2} is the attempt to write just as in the collinear case a single correlator $\Phi^{[-]}(x_1,p_{1\st}) = \tr_c\big[\Phi(x_1,p_{1\st})U_-^\dagger[p_1]U_-[p_1]\big]$. A complication is that in a TMD correlator the gauge link is a staple-like one running through minus infinity, i.e.\ $U_{[0,\xi]}=U_{[0,-\infty]}^nU_{[0_\st,\xi_\st]}^\st U_{[-\infty,\xi]}^n$. This complication prevents color-separation, which according to us is an integral part of the full treatment of soft and collinear gluons. It appears already as part of including gauge links at tree-level and does not necessarily contradict the treatment to incorporate soft factors in the TMD definition~\cite{Collins:2011zzd,Collins:2012uy}.

\section{Azimuthal expansion of the parton correlators}
In the above, both the TMD distribution functions $\Phi$ and the Drell-Yan cross section can be expanded in transverse moments, yielding
\bea
\Phi(x,p_\st) &=&\sum_m \Phi^{(m)}(x,p_\st^2)\,p_\st^m (\varphi), \\
d\sigma_{\text{DY}}(x_1,x_2,q_\st) 
&=&\sum_{m} d\sigma_{\text{DY}}^{(m)}(x_1,x_2,q_\st^2)\,q_\st^m (\varphi),
\eea
where the angle $\varphi$ represents the angular dependence of the transverse vectors $p_\st$ or $q_\st$, respectively and $p_\st^m(\varphi)$ is the symmetric traceless rank $m$ tensor constructed from the transverse momenta, i.e.\ 
\be 
p_\st^{\alpha_1\ldots\alpha_m} = p_\st^{\alpha_1}\ldots p_\st^{\alpha_m} - {\rm traces} 
\ \Longleftrightarrow\ \frac{\vert p_\st\vert^m}{2^{m-1}}\,e^{\pm im\varphi}.
\ee

By inverting these expressions, one can relate the definite rank TMDs $\Phi^{(m)}(x,p_\st^2)$ to the azimuthally integrated full TMD PDFs $\Phi(x,p_\st)$ weighted with $p_\st^m(\varphi)$, as explained in detail in Refs.~\cite{Buffing:2011mj,Buffing:2012sz}. The definite rank functions appearing in the expansion for $\Phi$ are actually quark or gluon correlators with in the matrix elements additional derivatives or gluonic fields, depending on the inserted operator being $iD_\st^\alpha$ or $A_\st^\alpha$ denoted as $\Phi_D^\alpha$, $\Phi_A^\alpha$, $\Phi_{DD}^{\alpha\beta}$, etc. In the treatment of TMD PDFs one needs actually only particular combinations of these correlators. Performing the transverse momentum weightings is sensitive to the nonlocality of the operators, in particular also the gauge links and their path. For example, for a TMD correlator with a gauge link $U$ one finds 
\bea
\Phi_\partial^{\alpha\,[U]}(x) & = &
\int d^2p_\st \ p_\st^\alpha\,\Phi^{[U]}(x,p_\st) 
\nonumber \\ &=& \widetilde\Phi_\partial^\alpha(x) + C_G^{[U]}\,\Phi_G^\alpha(x),
\eea
where $\widetilde\Phi_\partial^{\alpha}(x) = \Phi_D^\alpha(x)-\Phi_A^\alpha(x)$ is the difference between a quark correlator including a covariant derivative and the quark-gluon-quark correlator, while $\Phi_G^\alpha(x)$ is a gluonic pole matrix element, corresponding to the emission of a collinear gluon of zero momentum~\cite{Efremov:1981sh}. These functions are collinear and independent of the gauge link. That dependence is only in the gluonic pole coefficient $C_G^{[U]}$, see Ref.~\cite{Boer:2003cm}. For the simple staple gauge links $U_\pm$ the gluonic pole coefficients are $C_G^{[\pm]} = \pm 1$. Similarly, we have higher moments,
\bea
\Phi_{\partial\partial}^{\alpha\beta\,[U]}(x) & = &
\widetilde\Phi_{\partial\partial}^{\alpha\beta}(x) 
+ C_G^{[U]}\,\widetilde\Phi_{\{\partial G\}}^{\alpha\beta}(x)
\nonumber \\&&
\mbox{} + C_{GG,c}^{[U]}\,\Phi_{GG,c}^{\alpha\beta}(x),
\label{e:Phidd}
\eea 
etc. An extra index $c$ is needed if there are multiple possibilities to construct a color singlet as is the case for a field combination $\overline\psi\,GG\,\psi$, namely $\tr_c[GG\psi\overline\psi]$ ($c=1$) and $\tr_c[GG]\,\tr_c[\psi\overline\psi]/N_c$ ($c=2$). For the staple like links only one configuration is relevant, having $C_{GG,1}^{[\pm]} = 1$ and $C_{GG,2}^{[\pm]} = 0$, see Ref.~\cite{Buffing:2012sz}. The weighted results also allow a unique parametrization of the gauge link dependent TMD correlators in terms of a finite set of definite rank TMDs depending on $x$ and $p_\st^2$, azimuthal tensors and gluonic pole factors~\cite{Buffing:2012sz},
\bea
&&\Phi^{[U]}(x,p_\st) = \nonumber\\
&&\mbox{}\qquad\Phi(x,p_\st^2) 
+ \frac{p_{\st i}}{M}\,\widetilde\Phi_\partial^{i}(x,p_\st^2)
+ \frac{p_{\st ij}}{M^2}\,\widetilde\Phi_{\partial\partial}^{ij}(x,p_\st^2)
\nonumber \\ &&\mbox{}\qquad +
C_{G}^{[U]}\left\lgroup\frac{p_{\st i}}{M}\,\Phi_{G}^{i}(x,p_\st^2)
+ \frac{p_{\st ij}}{M^2}\,\widetilde\Phi_{\{\partial G\}}^{\,ij}(x,p_\st^2)
\right\rgroup
\nonumber \\ &&\mbox{}\qquad +
\sum_c C_{GG,c}^{[U]}\frac{p_{\st ij}}{M^2}\,\Gamma_{GG,c}^{ij}(x,p_\st^2).
\label{e:TMDstructure}
\eea
Depending on partons (quarks or gluons) and target, there is a maximum rank, which for quarks in a nucleon is rank 2. For gluons in a nucleon one has to go up to rank 3. Actually for the highest rank, time-reversal symmetry does not allow a time-reversal odd rank 2 correlator, i.e.\ $\widetilde\Phi_{\{\partial G\}} = 0$. Note that since the tensors $p_\st^{ij}$ on the rhs of Eq.~\ref{e:TMDstructure} are traceless and symmetric, the correlators they multiply also must be made traceless in order to make the identification of the correlators unique.

\section{Azimuthal expansion for the cross section}
In this situation, the weighting has to be done with the tensors $q_\st^{\alpha_1\ldots\alpha_m}$, which is in principle straightforward as $q_\st = p_{1\st} + p_{2\st}$, and thus involves a sum over various weightings. One gets at rank 2 among others contributions
\be
\langle p_{1\st}^{\alpha}\,p_{1\st}^{\beta}\,\sigma_{DY}\rangle = 
\frac{1}{N_c}\,\Phi_{\partial\partial}^{[-]\alpha\beta}(x_1)\,\Gamma^\ast\,
\overline\Phi{}^{[-^\dagger]}(x_2) \,\Gamma
\ee
and similarly for $\langle p_{2\st}^{\alpha}\,p_{2\st}^{\beta}\,\sigma_{DY}\rangle$, with $\Phi_{\partial\partial}^{[-]\alpha\beta}$ expanded as in Eq.~\ref{e:Phidd}. For the mixed contribution an additional complication arises because of the color entanglement of Wilson lines. Starting with Eq.~\ref{e:basic-2} one finds the expanded expression
\bea
\langle p_{1\st}^{\alpha}\,p_{2\st}^{\beta}\sigma_{DY}\rangle =\hspace{3mm}
\frac{1}{N_c}&&\hspace{-3mm}\widetilde\Phi_{\partial}^{\alpha}(x_1)\hspace{0.4mm}\,\Gamma^\ast\,\widetilde{\overline\Phi}{}_{\partial}^{\beta}(x_2)\hspace{0.4mm}\,\Gamma \nonumber \\
- \frac{1}{N_c}&&\hspace{-3mm}\Phi_{G}^{\alpha}(x_1)\,\Gamma^\ast\,\widetilde{\overline\Phi}{}_{\partial}^{\beta}(x_2)\hspace{0.4mm}\,\Gamma \nonumber \\
- \frac{1}{N_c}&&\hspace{-3mm}\widetilde\Phi_{\partial}^{\alpha}(x_1)\hspace{0.4mm}\,\Gamma^\ast\,\overline\Phi{}_{G}^{\beta}(x_2)\,\Gamma \nonumber \\
- \frac{1}{N_c^2-1}\frac{1}{N_c}&&\hspace{-3mm}\Phi_{G}^{\alpha}(x_1)\,\Gamma^\ast\,\overline\Phi{}_{G}^{\beta}(x_2)\,\Gamma.
\label{e:DYweight11}
\eea
To understand the prefactors one has to realize that the gluonic pole correlator $\Phi_G^\alpha(x_1)$ comes from a derivative acting on the gauge connection $U_-[p_1]$. This leads to a gluon field inserted in the correlator $\Phi(p_1)$ and a color charge $T^a$ at the position of the gauge connection in Fig~\ref{f:DYGL}. For the second and third term this does not lead to a different color factor as compared to the terms without azimuthal dependence, one just has the $1/N_c$ of the splitting of color traces as in Eq.~\ref{e:basic-0}, though a trace containing color charges arising from the gauge connections has to be included, giving
\[
\frac{\tr_c[T^aT^a]}{\tr_c[T^aT^a]\,\tr_c[1]} = \frac{1}{N_c}.
\]
The minus signs in the second and third term in Eq.~\ref{e:DYweight11} come from the gluonic pole factor multiplying the $\Phi_G$ correlators. For the last term with two gluonic pole correlators, the color factor is
\[
\frac{{\rm Tr}_c[T^aT^bT^aT^b]}{{\rm Tr}_c[T^aT^a]\,{\rm Tr}_c[T^bT^b]} = -\frac{1}{N_c^2-1}\frac{1}{N_c}.
\]
This implies not only a suppression of the asymmetry, but a sign change compared to naive parton calculations as well. In general, for higher weightings, the color factor is given by a ratio of symmetrized color charges,
\be
\frac{{\rm Tr}_c[T^{\{a_1}\hspace{-0.2mm}... T^{a_n\}}\hspace{-0.2mm}T^{\{b_1}\hspace{-0.2mm}... T^{b_m\}}\hspace{-0.2mm}T^{\{a_1}\hspace{-0.2mm}... T^{a_n\}}\hspace{-0.2mm}T^{\{b_1}\hspace{-0.2mm}... T^{b_m\}}\hspace{-0.2mm}]}
{{\rm Tr}_c[T^{\{a_1}\hspace{-0.2mm}... T^{a_n\}}\hspace{-0.2mm}T^{\{a_1}\hspace{-0.2mm}... T^{a_n\}}\hspace{-0.2mm}]\,
{\rm Tr}_c[T^{\{b_1}\hspace{-0.2mm}... T^{b_m\}}\hspace{-0.2mm} T^{\{b_1}\hspace{-0.2mm}... T^{b_m\}}\hspace{-0.2mm}]}. \nonumber
\ee
For instance, the result of a weighting with $p_{1\st}^{\alpha_1}\,p_{1\st}^{\alpha_2}\,p_{2\st}^{\beta}$ contains, among others, a term
\be 
- \,\tfrac{N_c^2+2}{(N_c^2-2)(N_c^2-1)}\,\tfrac{1}{N_c}\,\Phi_{GG}^{\alpha_1 \alpha_2}(x_1)\,\Gamma^\ast\,\overline{\Phi}{}_{G}^{\beta}(x_2)\,\Gamma ,
\ee
where the minus sign originates from a gluonic pole coefficient and a weighting with $p_{1\st}^{\alpha_1}\,p_{1\st}^{\alpha_2}\,p_{2\st}^{\beta_1}\,p_{2\st}^{\beta_2}$ contains
\be 
+ \,\tfrac{3N_c^4-8N_c^2-4}{(N_c^2-2)^2 (N_c^2-1)}\,\tfrac{1}{N_c}\,\Phi_{GG}^{\alpha_1 \alpha_2}(x_1)\,\Gamma^\ast\,\overline{\Phi}{}_{GG}^{\beta_1 \beta_2}(x_2)\,\Gamma.
\label{e:DYweight220}
\ee
These color factors~\cite{Sjodahl:2012nk} depend on the gluonic rank only and are insensitive to the presence of partial derivative terms. They in general imply a suppression of multiple gluonic pole contributions in azimuthal asymmetries.

The resulting tree level cross section for the Drell-Yan process at measured $q_\st$ thus can be written as a sum of various contributions, each having their characteristic azimuthal dependence. There is color entanglement for gluonic pole contributions, but it is still possible to write a factorized expression for each term after inclusion of the full gauge links (resummed to all orders). For a given harmonic $\varphi$-dependence in $q_\st$, there will not only be a split up in various terms depending on polarizations of hadrons and partons, but there will also be a dependence on the gluonic rank of the functions with process dependent color factors. Formulated slightly more general for hadron-hadron scattering, each contribution in the squared amplitude is a convolution in transverse momentum ($q_\st = p_{1\st} + p_{2\st}$), but is also assigned an additional color factor beyond the basic $1/N_c$. Omitting the $Q$-dependence, one finds
\bea
\sigma(x_1,x_2,q_\st) &\sim & \frac{1}{N_c}f_{R_{G1}R_{G2}}^{[U_1,U_2]}
\,\Phi^{[U_1]}(x_1,p_{1\st}) 
\nonumber \\
&& \mbox{}
\otimes \overline{\Phi}^{[U_2]}(x_2,p_{2\st})
\,\hat\sigma (x_1,x_2),
\eea
where the correlators are expanded as in Eq.~\ref{e:TMDstructure}, including the gluonic pole factors. The process dependent factors $f_{R_{G1}R_{G2}}^{[U_1,U_2]}$ depend on the gluonic pole ranks $R_{G1}$ and $R_{G2}$ of the contributing terms in the hadronic correlators as well as on the gauge link structures of both TMDs. For Drell-Yan, all relevant factors for quark correlators have been tabulated in Table~\ref{t:DYfactors}. For processes more complicated than Drell-Yan a dependence on the color flow possibilities of both TMDs will appear, as mentioned before. This will be explained in more detail in a future paper. As explained in Ref.~\cite{Buffing:2013kca}, for gluons the expansion of the gauge link dependent correlators goes up to rank 3. 

\begin{table}[!tb]
\centering
\begin{tabular}{|>{\centering}m{19mm}">{\centering}m{20.1mm}|>{\centering}m{20.1mm}|>{\centering}m{20.1mm}|}
\hline
&\multicolumn{3}{c|}{$R_G$ for $\Phi^{[-]}$} \tabularnewline[2pt]
$R_G$ for $\overline{\Phi}^{[-^\dagger]}$	& $0$		& $1$	& \vspace{2pt} $2$ \vspace{-2pt}
\tabularnewline \thickline
0	& $1$	& $1$															& $1$
\tabularnewline[4pt]
1	&	$1$	& $-\tfrac{1}{N_c^2-1}$								& $\tfrac{N_c^2+2}{(N_c^2-2)(N_c^2-1)}$
\tabularnewline[4pt]
2	&	$1$	& $\tfrac{N_c^2+2}{(N_c^2-2)(N_c^2-1)}$	& $\tfrac{3N_c^4-8N_c^2-4}{(N_c^2-2)^2 (N_c^2-1)}$
\tabularnewline[4pt] \hline
\end{tabular}
\parbox{0.45\textwidth}{\caption{\textit{The factor $f_{R_{G1}R_{G2}}^{[-,-^\dagger]}$ as a function of the gluonic pole ranks of both the quark and antiquark correlator in the Drell-Yan process.}}
\label{t:DYfactors}}
\end{table}

\section{Discussion and conclusions}
As discussed in Ref.~\cite{Buffing:2012sz}, there are only two TMDs with a gluonic pole rank 1 that contribute at leading order, the Boer-Mulders function $h_1^{\perp}$ and the Sivers function $f_{1T}^{\perp}$. At (total) rank 2, there are three universal Pretzelocity functions, $h_{1T}^{\perp(A)}$ with gluonic pole rank 0 and $h_{1T}^{\perp(B1)}$ and $h_{1T}^{\perp(B2)}$ with gluonic pole rank 2. Considering different processes one finds in general linear combinations that can be used to isolate the universal functions. All the other TMDs that are relevant at leading order have a gluonic pole rank of 0. This implies that the first azimuthal asymmetries~\cite{Tangerman:1994eh} where addional color factors appear are the double Sivers asymmetry, which requires polarized beams and the double Boer-Mulders asymmetry, which is accessible using unpolarized hadron beams only~\cite{Ralston:1979ys}, such as at the LHC. 

One might wonder why the gauge connections cannot be disentangled as in the collinear case, since the color charges are entangled in both cases. An attempt in this direction was made in Ref.~\cite{Dunnen:2013thesis}. The difference between the collinear and the TMD case, however, is that multiple directions are involved, a light-like direction $n$ in the Sudakov expansion of the momenta and transverse directions. These transverse directions are different for the gauge connections labeled with $p_1$ and $p_2$. For the light-like direction $n$ one can at leading order in principle make {\em one} choice for the full process using the fact that varying $n$ is $1/Q$ suppressed. Having just one direction in the gauge connections, it is straightforward to see that the entanglement of gauge connections can be undone by a gauge transformation, but complications arise if there are multiple directions. 

It should also be noted that the results in this paper do not prove factorization. We have used the tree-level result for Drell-Yan and shown that the inclusion of gauge links gives an additional color prefactor. Additional effects, like the inclusion of next-to-leading order contributions or factorization breaking effects have not been taken into account. For the latter we refer to Ref.~\cite{Collins:2008sg} for the case of hadroproduction in hadron-hadron scattering. Also the inclusion of soft factors in the TMD definition is still needed, possibly with some modifications~\cite{Collins:2011zzd} or using techniques from soft collinear effective theory (SCET)~\cite{GarciaEchevarria:2011rb}. In Refs.~\cite{Avsar:2012hj} and \cite{Dominguez:2010xd} issues on gluon TMD factorization in the low-x regime and diffraction are addressed.
\bigskip

Summarizing, for hadroproduction processes gauge links lead to color entanglement, which is already the case in the tree-level description of the Drell-Yan process at measured $q_\st$. Factorized expressions can still be obtained, however, but they deviate from the naive expectation. It turns out that additional color factors have to be included in the expressions for azimuthal asymmetries. Due to these additional color factors, asymmetries are suppressed for Drell-Yan and additional sign changes appear.
\bigskip

\begin{acknowledgments}
This research is part of the research program of the ``Stichting voor Fundamenteel Onderzoek der Materie (FOM)", which is financially supported by the ``Nederlandse Organisatie voor Wetenschappelijk Onderzoek (NWO)". We also acknowledge support of the FP7 EU-programmes HadronPhysics3 (contract no 283286) and QWORK (contract 320389). We would like to acknowledge discussions with Wilco den Dunnen, Asmita Mukherjee and Wouter Waalewijn. All figures were made using JaxoDraw~\cite{Binosi:2003yf}.
\end{acknowledgments}

\end{document}